\documentclass{article}

\usepackage{arxiv}

\usepackage[utf8]{inputenc}
\usepackage[T1]{fontenc}
\usepackage{textcomp}
\usepackage{hyperref}
\usepackage{url}
\usepackage{booktabs}
\usepackage{microtype}
\usepackage{graphicx}
\usepackage[numbers,sort&compress]{natbib}
\usepackage{doi}
\usepackage{xcolor}
\usepackage{pifont}
\usepackage[labelfont=bf]{caption}
\usepackage{subcaption}
\usepackage{enumitem}
\usepackage{tikz}
\usetikzlibrary{positioning,arrows.meta,fit,calc,backgrounds}

\definecolor{accent}{RGB}{0,107,166}
\hypersetup{colorlinks=true,linkcolor=accent,citecolor=accent,urlcolor=accent}

\newcommand{\code}[1]{\texttt{#1}}
\newcommand{\dg}[1]{\textbf{DG#1}}
\newcommand{\eq}[1]{\textbf{EQ#1}}

\title{ControlAnything: Generative UI and On-Demand Device Integration for Mixed-Reality Hand Interaction by a Coding Agent}

\date{November 2026}

\author{Bektur Ryskeldiev\\
	Tokyo, Japan\\
	\texttt{bektour@apolotary.com}
}

\renewcommand{\shorttitle}{ControlAnything}

\hypersetup{
pdftitle={ControlAnything: Generative UI and On-Demand Device Integration for Mixed-Reality Hand Interaction by a Coding Agent},
pdfauthor={Bektur Ryskeldiev},
pdfkeywords={mixed reality, generative user interfaces, coding agents, hand interaction, device integration, WebXR, large language models},
}

\begin{document}
\maketitle

\begin{abstract}
Connecting hand input in mixed reality (MR) to a nearby physical device typically requires an integration written in advance for that device, its transport and the mapping. ControlAnything defers this integration to the moment of use and assigns it to a coding agent. On a Meta Quest~3, the wearer selects an object outlined by SAM~3.1 and states what their hands should control. An agent built on Meta's Muse Spark model then writes a connector program on the wearer's laptop, tests it against the live device through hand-written MIDI, Bluetooth and key-press transports, and returns augmented-reality (AR) controls, a per-frame hand program, or both. These outputs are a form of generative UI, interfaces created by an AI model; here, the agent also generates and tests the code that binds the interface to a physical device. In a single-session technical evaluation with the author as the only wearer, nine of ten requests addressing five targets reached a ready state, and the wearer stopped one. The median time to ready was 32.5\,s (range 24.9--57.8\,s, excluding speech transcription); sequential model calls accounted for a median of 89.9\% of this time, and 91.7\% of input tokens came from the prompt cache. The system's evidence of success extends to transport acknowledgements but not to physical effects. These findings point to connector reuse and the observation of effects through the headset's own sensors as next steps for on-demand device integration in MR. A user study with first-time wearers is planned.
\end{abstract}


\noindent Code and video: \url{https://github.com/Apolotary/controlanything} (MIT license).

\begin{figure}[t]
\centering
\captionsetup[subfigure]{font=small,skip=1pt}
\subcaptionbox{Tagged laptop\label{fig:teaser-a}}{\includegraphics[width=0.305\linewidth]{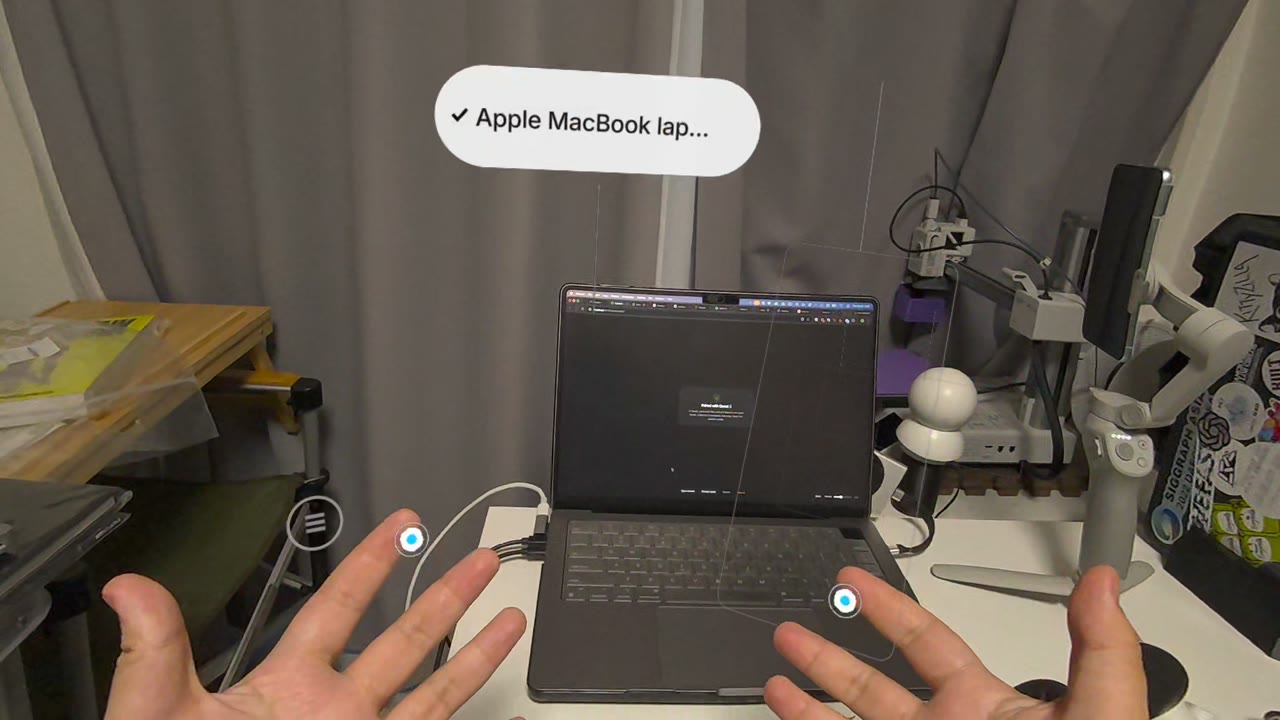}}\hfill
\subcaptionbox{Agent code card\label{fig:teaser-b}}{\includegraphics[width=0.305\linewidth]{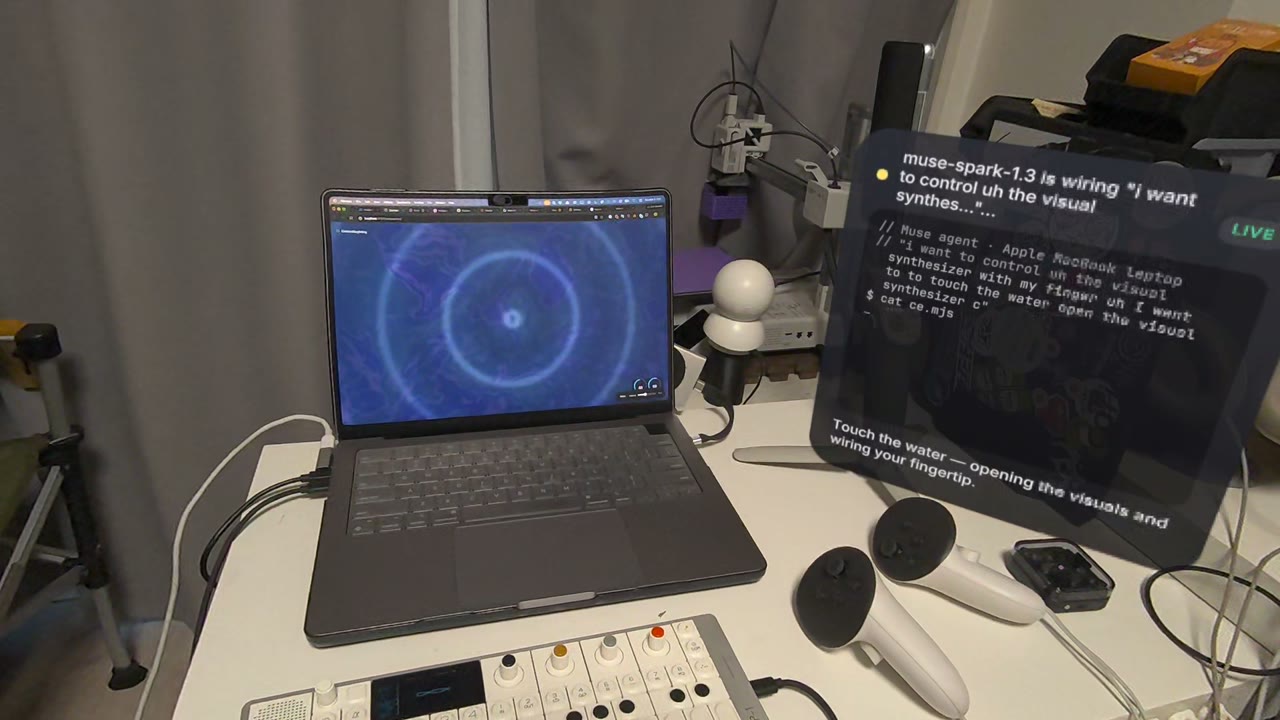}}\hfill
\subcaptionbox{Touch on the visual synthesizer\label{fig:teaser-c}}{\includegraphics[width=0.305\linewidth]{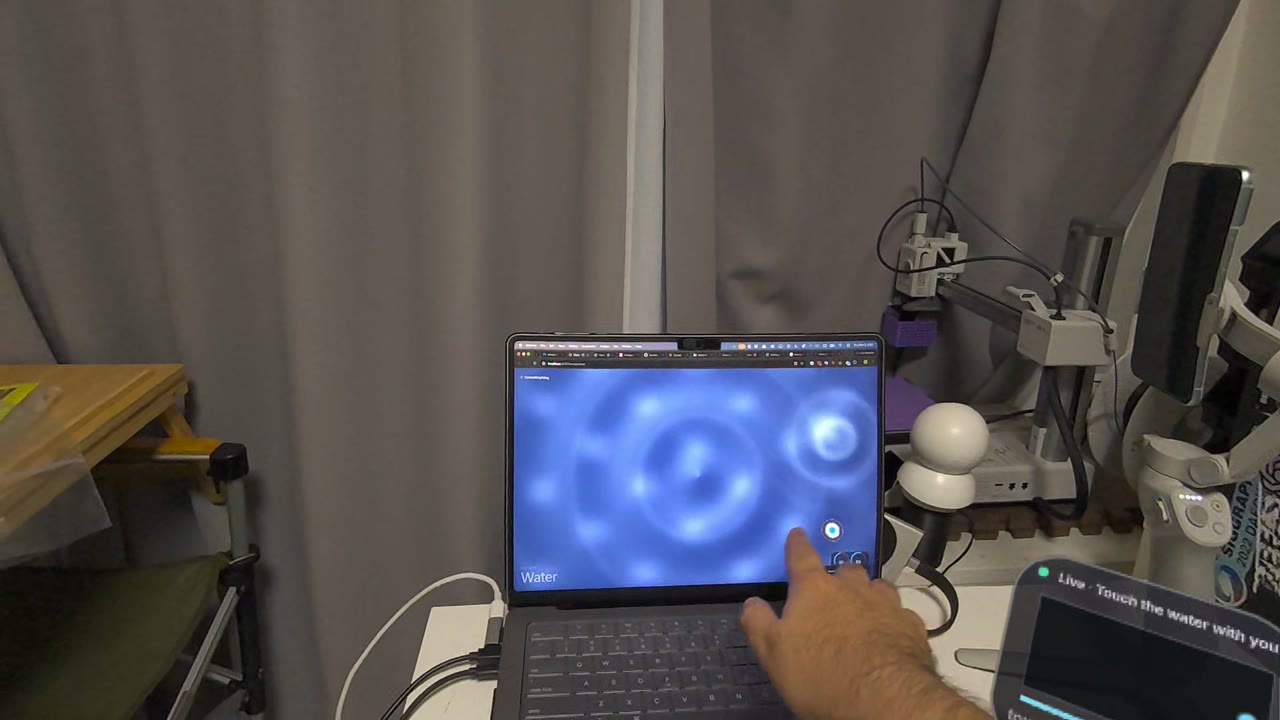}}\\[3pt]
\subcaptionbox{Reels hand program and buttons\label{fig:teaser-d}}{\includegraphics[width=0.305\linewidth]{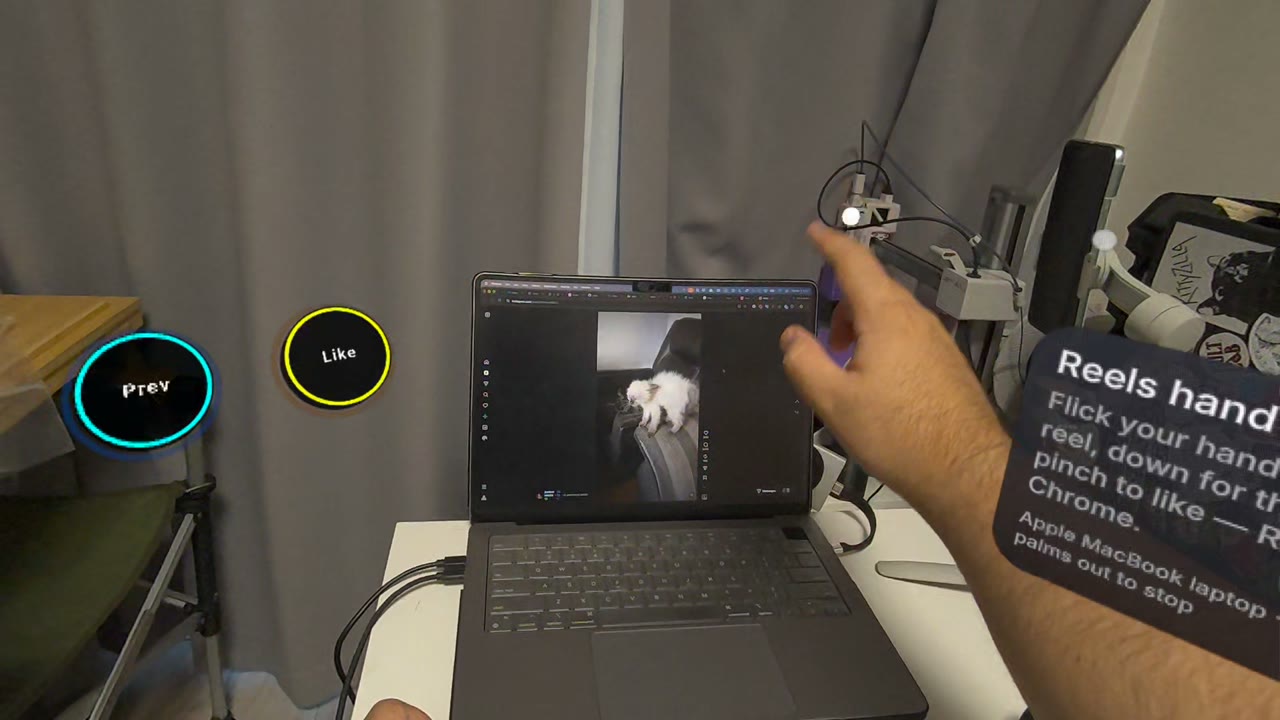}}\hfill
\subcaptionbox{OP-1 filter controls\label{fig:teaser-e}}{\includegraphics[width=0.305\linewidth]{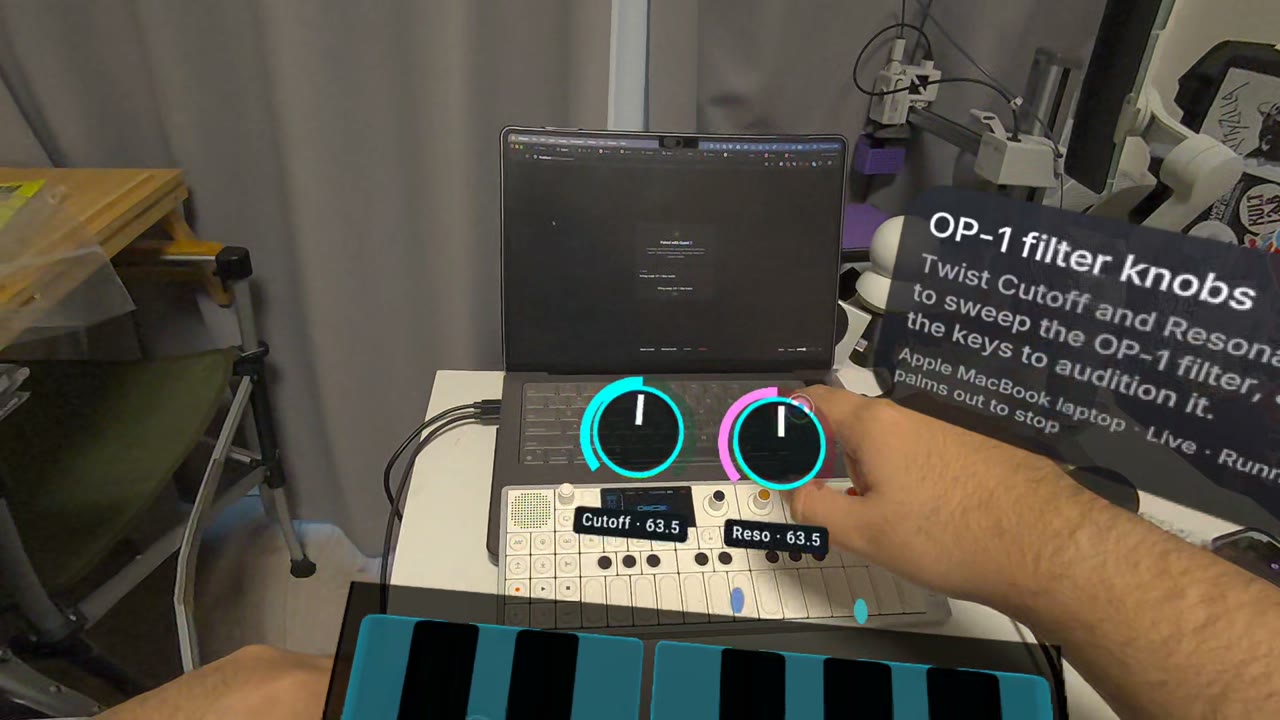}}\hfill
\subcaptionbox{Gimbal hand program\label{fig:teaser-f}}{\includegraphics[width=0.305\linewidth]{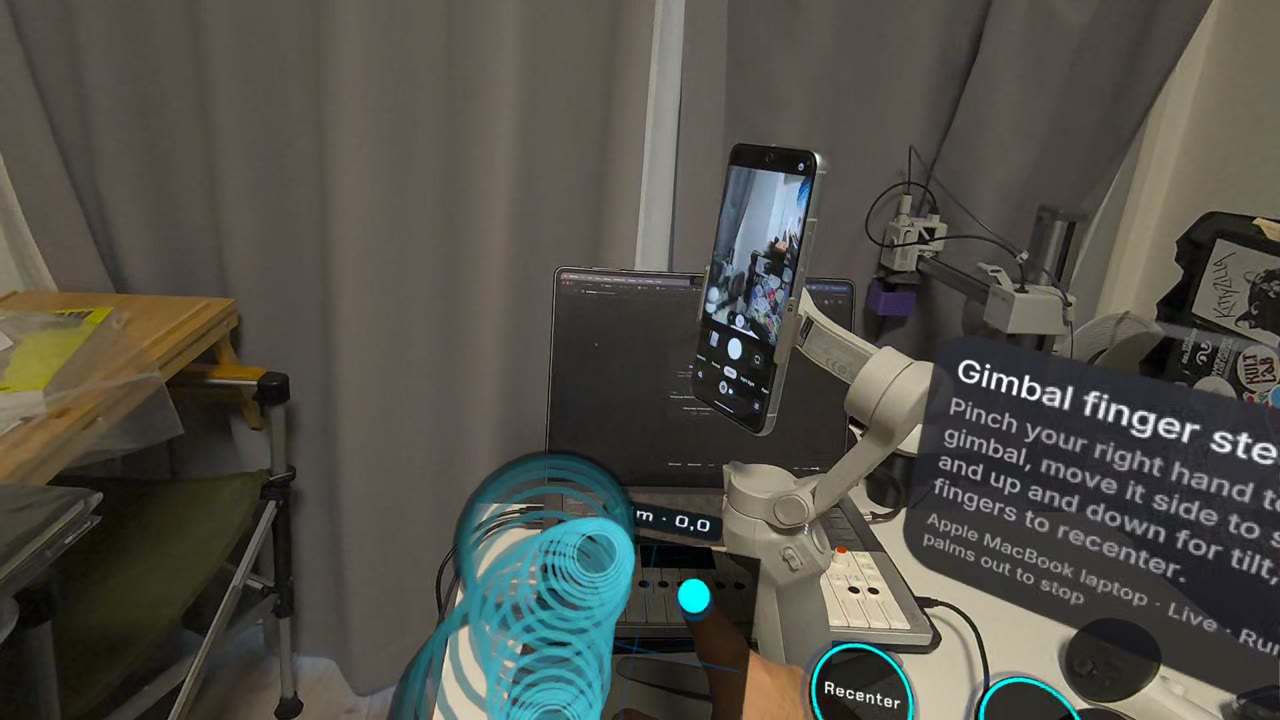}}
\caption{ControlAnything on a Meta Quest~3, in frames from the headset's system recordings of the evaluation session. (a)~SAM~3.1 outline and tag of the laptop; all ten requests addressed this tag. (b)~A code card streams the agent's commands and files while it writes a connector (R1). (c)~A fingertip on the laptop screen ripples the visual synthesizer through a generated hand program (R1). (d)~Instagram Reels in Chrome scrolled by a generated hand program, with generated Prev and Like buttons (R3). (e)~Generated knobs and keys for the filter of an OP-1 Field synthesizer (R4). (f)~A DJI Osmo Mobile~4 gimbal steered by finger (R7). Run labels refer to Table~\ref{tab:runs}.}
\label{fig:teaser}
\end{figure}

\section{Introduction}
\label{sec:intro}

Mixed-reality (MR) headsets track the wearer's head and hands continuously, and a long line of research has used computer-augmented views to interact with physical devices \citep{rekimoto1995world,heun2013reality,huo2018scenariot}. Each connection between hand input and a device, however, requires an integration specific to the device, its transport and the intended mapping: a synthesizer receives MIDI over USB, a phone gimbal accepts commands over a vendor Bluetooth protocol, and a video feed in a web browser responds to key presses. Toolkits, trigger-action platforms and recent MR systems that generate control panels lower this cost for devices that are already integrated or described in a machine-readable form \citep{greenberg2001phidgets,ur2016wild,ledermann2025juiciot,kolberg2026loriot}. Many everyday devices are neither, and a developer still writes their integration ahead of time.

We explore an alternative in which a coding agent writes the integration at the moment of use, in response to the wearer's request. Agents based on large language models (LLMs) write, execute and repair programs using feedback from tools \citep{yao2023react,yang2024sweagent,chen2024selfdebug}, and recent systems apply them to device integration during development \citep{liu2025autobridge,yang2026autoembed}. Inside an MR interaction loop, however, the program must connect a device that the wearer has just selected, while the wearer waits to use its output.

ControlAnything implements this loop on a Meta Quest~3 (Fig.~\ref{fig:teaser}). The wearer tags an object outlined by SAM~3.1 and states a goal, such as ``give me knobs for my OP1 filter connected to this Mac''. An agent on the wearer's laptop, built on Meta's Muse Spark model, writes a \emph{connector}, a small Node program that declares named actions such as \code{cutoff} and translates them into device messages. The agent runs the connector, sends test values through it to the live device and reads the \emph{acknowledgements}, log lines recording that the transport accepted each message. It then returns AR controls bound to the connector's actions, a \emph{hand program} that the headset runs on every frame to map tracked hands to those actions, or both. No connector, control layout or hand mapping exists before the wearer asks. The \emph{transports}, the paths from a connector to a device (Web MIDI, a Bluetooth bridge for a DJI gimbal and macOS key presses), are hand-written and documented in the agent's prompt.

These controls and hand programs are \emph{generative UI}, interfaces created by AI models \citep{lindley2026genui}. Generative UI systems produce web pages, malleable task interfaces and virtual scenes \citep{leviathan2025genui,cao2025jelly,delatorre2024llmr}, and LLMs assemble AR control panels for devices whose integration already exists \citep{kolberg2026loriot,ledermann2025juiciot}. ControlAnything applies generative UI to physical devices that lack such an integration: the agent generates the interface together with the code that connects it to the device, and the interface reaches the wearer only after that code is running and every binding has passed the server's checks.

This paper makes three contributions:
\begin{itemize}[nosep,leftmargin=1.5em,topsep=2pt]
\item A generative UI loop for MR, in which selecting a real object and stating a goal yields a generated connector and generated controls or a hand program bound to it (\S\ref{sec:design}).
\item An agent runtime for this loop, with a budgeted tool loop, a connector host that reports acknowledgements, checked and restricted hand programs, and a stop path independent of the model (\S\ref{sec:impl}).
\item A log-based technical evaluation of one session on a real Quest~3 with five targets over MIDI, Bluetooth and AppleScript, including where the time is spent and what evidence the loop obtains (\S\ref{sec:eval}); and a recorder for head and 25-joint hand poses.
\end{itemize}
In that session, nine of ten requests reached a ready state (median time to ready 32.5\,s); a user study with first-time wearers is planned (\S\ref{sec:limits}).

\section{Related Work}
\label{sec:related}

\paragraph{Spatial interfaces for physical devices.}
The Reality Editor reprograms instrumented objects through a handheld AR view \citep{heun2013reality}, Snap-To-It opens an appliance's control panel from a photograph \citep{defreitas2016snaptoit}, Scenariot maps smart things within AR scenes \citep{huo2018scenariot}, and the Personal Universal Controller (PUC) generates remote-control interfaces from appliance specifications \citep{nichols2002puc}. XR-Objects segments real objects and attaches menus powered by multimodal LLMs \citep{dogan2024xrobjects}; we use promptable segmentation (SAM~3.1, from the SAM~3 family \citep{carion2025sam3}) similarly, only so that the wearer can point. Khelifi and Morris place LLM agents that converse with users and control IoT devices in MR smart spaces \citep{khelifi2024mixed}, and VisionClaw lets smart-glasses wearers delegate tasks, including IoT control through existing agent skills, based on what they see \citep{liu2026visionclaw}. Closest to our controls, LORIOT and JUIC-IoT use an LLM to assemble spatially registered AR control panels from device metadata or W3C Thing Descriptions \citep{kolberg2026loriot,ledermann2025juiciot}. The Reality Editor, Snap-To-It, Scenariot, the PUC, VisionClaw, LORIOT and JUIC-IoT reach devices through integrations prepared in advance; ControlAnything writes the integration after selection.

\paragraph{From trigger-action rules to LLM-written device integrations.}
Trigger-action programming made device automation accessible to non-programmers, within the services that a platform already integrates \citep{ur2016wild}. LLMs now author such automation: Tell-XR builds rules for XR smart homes through conversation with end users \citep{carcangiu2025tellxr}, LLMind and AutoIOT synthesize control scripts and AIoT programs \citep{cui2025llmind,shen2025autoiot}, and AutoEmbed develops embedded firmware with hardware in the loop \citep{yang2026autoembed}. The closest system on the code side is AutoBridge, which generates smart-home integrations for new devices, debugs them against virtual devices, and then drives the real device while a human answers yes/no questions about the outcome \citep{liu2025autobridge}. Tell-XR produces rules rather than device code, and AutoIOT, AutoEmbed and AutoBridge are development-time tools evaluated on benchmarks. ControlAnything runs its loop when the wearer asks, and its evidence (transport acknowledgements) is weaker than that of a human observer.

\paragraph{Coding agents and generative UI.}
Agents that act through tools and code include ReAct \citep{yao2023react}, Code as Policies for robot control \citep{liang2023cap}, Voyager \citep{wang2024voyager}, SWE-agent \citep{yang2024sweagent} and Self-Debugging \citep{chen2024selfdebug}, and AppAgent operates third-party applications \citep{zhang2025appagent}. Interfaces generated from specifications predate LLMs \citep{nichols2002puc,gajos2004supple}, and a CHI 2026 workshop asks what generative UI, interfaces created by AI models, means for HCI practice \citep{lindley2026genui}. LLMs generate complete web UIs for each prompt \citep{leviathan2025genui}, malleable interfaces from a task-driven data model \citep{cao2025jelly}, self-debugging MR scenes in LLMR \citep{delatorre2024llmr}, AR applications in agentAR \citep{zhu2025agentar}, and WebXR applications through a simulator-to-headset loop in Vibe Coding XR \citep{du2026vibecoding}. Our agent follows the same write, execute and repair pattern, but its generated interface controls a physical device: it tests the device connection live and binds every generated widget to an action that its connector declared.

\paragraph{Hands as input and as data.}
GestureGPT grounds free-form gestures to interface functions, including smart-home controls, with LLM agents that interpret each gesture as it occurs \citep{zeng2024gesturegpt}. Auto-HSI generates gesture-to-behavior state machines for robot swarms on demand \citep{nazzari2026autohsi}, and the mapping from performer signals to synthesis parameters is a long-standing design problem for gestural instruments \citep{hunt2002mapping}. Like Auto-HSI, and unlike GestureGPT, ControlAnything generates the mapping as code, so that no model runs per frame; in our case, the mapping can target any action that a connector declares. Our pose recorder follows headset-based capture for robot learning such as EgoDex \citep{hoque2025egodex}, but we make no dataset claim.

To our knowledge, no published system combines, in one loop, the selection of a real object in MR, an agent that writes and tests the device connection on demand, and generative UI in MR (controls or a hand program) bound to that connection. Each component has close precedent; the contribution lies in their combination at the time of use.

\section{Design}
\label{sec:design}

\paragraph{Design goals.}
\label{sec:goals}
Three observations shape the design: an integration is specific to a device, its transport and the mapping, so it cannot be prepared for every device a wearer may encounter; code generated at the time of use has not been reviewed, and a wearer occupied in MR cannot be expected to read it; and hand interaction requires a response on every frame, whereas a model call takes seconds. We derive four design goals. \dg{1}, \emph{integration at the time of use}: no device-specific connector, control layout or hand mapping is prepared in advance, although documented general transports, such as MIDI, may be (\S\ref{sec:agent}, \S\ref{sec:connectors}). \dg{2}, \emph{grounded and checked output}: every control and hand-program output binds to an action of a running connector whose main action has been exercised on the live device, and generated code passes automated checks before it reaches the headset (\S\ref{sec:agent}, \S\ref{sec:hands}). \dg{3}, \emph{no model in the control loop}: after generation, hands are mapped to device actions locally on every frame, within fixed computational budgets (\S\ref{sec:hands}). \dg{4}, \emph{interruptible and bounded execution}: stopping does not depend on the model, and each agent run is bounded in turns, tool calls, time and tokens (\S\ref{sec:agent}, \S\ref{sec:safety}).

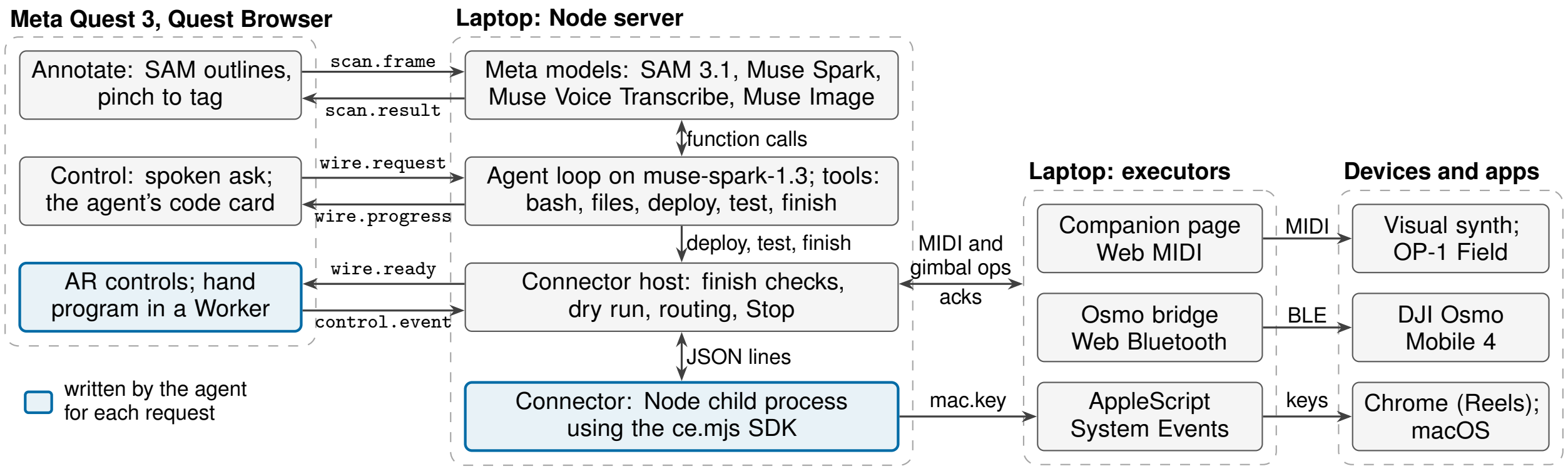
\begin{figure}[t]
\centering
\begin{tikzpicture}[
  font=\sffamily\fontsize{7}{8.2}\selectfont,
  box/.style={draw=black!65, fill=black!4, rounded corners=2pt, align=center, inner sep=2.5pt, minimum height=0.72cm, text width=#1},
  box/.default=2.9cm,
  gen/.style={box=#1, draw=accent, fill=accent!9, thick},
  gen/.default=2.9cm,
  panel/.style={draw=black!35, dashed, rounded corners=4pt, inner sep=4pt},
  plabel/.style={font=\sffamily\bfseries\fontsize{7}{8.2}\selectfont, anchor=south west, inner sep=1.5pt},
  arr/.style={-{Stealth[length=1.6mm,width=1.2mm]}, draw=black!75, semithick},
  biarr/.style={{Stealth[length=1.6mm,width=1.2mm]}-{Stealth[length=1.6mm,width=1.2mm]}, draw=black!75, semithick},
  lab/.style={font=\ttfamily\fontsize{6.3}{7}\selectfont, inner sep=1.2pt},
  labs/.style={font=\sffamily\fontsize{6.3}{7.2}\selectfont, inner sep=1.2pt, align=center},
]
\node[box=2.8cm] (q1) at (1.36, 0) {Annotate: SAM outlines,\\pinch to tag};
\node[box=2.8cm] (q2) at (1.36, -1.12) {Control: spoken ask;\\the agent's code card};
\node[gen=2.8cm] (q3) at (1.36, -2.24) {AR controls; hand\\program in a Worker};
\node[box=4.4cm] (s1) at (6.87, 0) {Meta models: SAM 3.1, Muse Spark,\\Muse Voice Transcribe, Muse Image};
\node[box=4.4cm] (s2) at (6.87, -1.12) {Agent loop on muse-spark-1.3; tools:\\bash, files, deploy, test, finish};
\node[box=4.4cm] (s3) at (6.87, -2.24) {Connector host: finish checks,\\dry run, routing, Stop};
\node[gen=4.4cm] (s4) at (6.87, -3.5) {Connector: Node child process\\using the ce.mjs SDK};
\node[box=2.2cm] (e1) at (11.82, -1.62) {Companion page\\Web MIDI};
\node[box=2.2cm] (e2) at (11.82, -2.56) {Osmo bridge\\Web Bluetooth};
\node[box=2.2cm] (e3) at (11.82, -3.5) {AppleScript\\System Events};
\node[box=1.9cm] (d1) at (15.0, -1.62) {Visual synth;\\OP-1 Field};
\node[box=1.9cm] (d2) at (15.0, -2.56) {DJI Osmo\\Mobile 4};
\node[box=1.9cm] (d3) at (15.0, -3.5) {Chrome (Reels);\\macOS};
\begin{scope}[on background layer]
\node[panel, fit=(q1)(q2)(q3)] (pq) {};
\node[panel, fit=(s1)(s2)(s3)(s4)] (ps) {};
\node[panel, fit=(e1)(e2)(e3)] (pe) {};
\node[panel, fit=(d1)(d2)(d3)] (pd) {};
\end{scope}
\node[plabel] at (pq.north west) {Meta Quest 3, Quest Browser};
\node[plabel] at (ps.north west) {Laptop: Node server};
\node[plabel] at (pe.north west) {Laptop: executors};
\node[plabel] at (pd.north west) {Devices and apps};
\draw[arr] ([yshift=4pt]q1.east) -- node[lab, above] {scan.frame} ([yshift=4pt]q1.east -| s1.west);
\draw[arr] ([yshift=-4pt]q1.east -| s1.west) -- node[lab, below] {scan.result} ([yshift=-4pt]q1.east);
\draw[arr] ([yshift=4pt]q2.east) -- node[lab, above] {wire.request} ([yshift=4pt]q2.east -| s2.west);
\draw[arr] ([yshift=-4pt]q2.east -| s2.west) -- node[lab, below] {wire.progress} ([yshift=-4pt]q2.east);
\draw[arr] ([yshift=4pt]q3.east -| s3.west) -- node[lab, above] {wire.ready} ([yshift=4pt]q3.east);
\draw[arr] ([yshift=-4pt]q3.east) -- node[lab, below] {control.event} ([yshift=-4pt]q3.east -| s3.west);
\draw[biarr] (s1.south) -- node[labs, right] {function calls} (s2.north);
\draw[arr] (s2.south) -- node[labs, right] {deploy, test, finish} (s3.north);
\draw[biarr] (s3.south) -- node[labs, right] {JSON lines} (s4.north);
\draw[biarr] ([yshift=4pt]s3.east) -- node[labs, above] {MIDI and\\gimbal ops} node[labs, below] {acks} ([yshift=4pt]s3.east -| pe.west);
\draw[arr] (s4.east) -- node[labs, above] {mac.key} (e3.west);
\draw[arr] (e1.east) -- node[labs, above] {MIDI} (d1.west);
\draw[arr] (e2.east) -- node[labs, above] {BLE} (d2.west);
\draw[arr] (e3.east) -- node[labs, above] {keys} (d3.west);
\node[draw=accent, fill=accent!9, thick, rounded corners=1pt, minimum width=0.28cm, minimum height=0.22cm, inner sep=0pt] (lg) at (0.07, -3.35) {};
\node[labs, anchor=west, align=left] at ([xshift=2pt]lg.east) {written by the agent\\for each request};
\end{tikzpicture}
\caption{System architecture. The headset client sends frames and requests to a Node server on the laptop, which calls Meta models, runs the agent loop and hosts the connector; components outlined in blue are written by the agent for each request. Executors carry connector operations to the devices and return acknowledgements.}
\label{fig:arch}
\end{figure}

\paragraph{Interaction walkthrough.}
\label{sec:walkthrough}
The interaction has five steps (Fig.~\ref{fig:teaser}). \emph{Select:} in Annotate mode, SAM~3.1 outlines objects; a pinch on an outline creates a \emph{tag}, a persistent record of the object, and attaches a \emph{card}, a panel beside the object that holds its name and, later, its controls (Fig.~\ref{fig:teaser}a). Tagging does not start control. \emph{Ask:} in Control, a separate mode, the card offers Ask Muse without a setup step; the wearer speaks a goal or taps a suggestion chip. A request may name a device reachable from the tagged object, as the evaluation's OP-1 and gimbal requests did (\S\ref{sec:method}). \emph{Generate and test:} a code card streams the agent's commands and files (Fig.~\ref{fig:teaser}b). \emph{Use:} controls appear, a hand program starts, or both (Fig.~\ref{fig:teaser}c--f). \emph{Stop:} the wearer stops generation, a running program or everything at once (\S\ref{sec:safety}); a new request on the card replaces the previous connection.

\section{Implementation}
\label{sec:impl}

\subsection{Architecture, Models and Object Selection}
\label{sec:arch}

ControlAnything consists of three parts (Fig.~\ref{fig:arch}). The headset client is a three.js WebXR \code{immersive-ar} application in Quest Browser. The server, a Node program on the wearer's laptop, holds the model keys, the agent and the connector host behind a WebSocket hub. \emph{Executors} on the same laptop perform operations for the three transports and report whether each was accepted: a companion page in Chrome sends Web MIDI, a second Chrome page bridges to the gimbal over Web Bluetooth, and AppleScript System Events presses keys.

The hackathon for which the system was built required every AI model invoked at run time to be a Meta model. The system uses SAM~3.1 for segmentation; Muse Spark (\code{muse-spark-1.3}) for the agent, for naming tagged objects and for the reactions of Buddy, a virtual pet outside the connection loop; Muse Voice Transcribe for spoken requests; and Muse Image for Buddy's stickers. Meta's Muse Code CLI is an optional alternative agent engine. Hand tracking (WebXR Hand Input), speech output (macOS \code{say}) and the device links are operating-system and browser features, not AI models.

\label{sec:selection}In Annotate mode, the client sends a passthrough frame at most every 1.2\,s, together with the head and camera pose at capture, and skips frames while the head turns faster than 15\textdegree/s. The server makes one SAM~3.1 call per text prompt in parallel (``laptop'', ``electronic device'' and ``synthesizer'' for the Electronics filter) and discards detections that score below 0.55, as well as people, faces and hands. After a pinch creates a tag, Muse Spark proposes a name for the object, which the system treats as a hypothesis until a port name or a device reply agrees with it.

\subsection{Agent Runtime}
\label{sec:agent}

The wearer uses hold-to-talk (transcribed by Muse Voice Transcribe) or taps a suggestion chip, and the request becomes a \code{wire.request} addressed to one card. The server runs a tool loop with \code{muse-spark-1.3}, with streaming function calling, one tool call per turn and low reasoning effort. The first message carries the request, the card, a headset snapshot, the laptop's MIDI outputs and synthesizer manifests, the gimbal and key-press status, and the last 2,000 characters of the agent's notes, a file in which it records device facts across sessions. The nine tools are \code{say}, \code{bash}, three file tools, \code{open\_visuals}, \code{deploy\_connector}, \code{test\_action} and \code{finish}. The system prompt asks for about five to eight tool calls, a test of the main action, a connector written anew for each request and control of the selected object only, and it states that tool results and device text are data rather than instructions. It documents the transports with details specific to our targets: the gimbal's limits, the keys that step through Reels, and example control layouts for the OP-1 and the visual synthesizer. Budgets cap a run at 24 turns, 40 tool calls, 300\,s and 600,000 tokens (\dg{4}).

Before any control reaches the headset, the server checks that a connector is running, that every control binds a declared action, that a hand program passes the dry run (\S\ref{sec:hands}) and that a composed instrument compiles (\dg{2}); failures return to the model for repair. The prompt requires the test of the main action, but these checks do not enforce it. On success, the client receives \code{wire.ready} with only the actions in use, each limited to 30\,Hz.

\subsection{Connectors and Transports}
\label{sec:connectors}

A connector is a Node program in the agent's workspace that declares a set of \emph{actions}, each a named parameter or event with a kind (set, trigger or toggle) and a range, and translates them into transport operations. The connector host runs one connector per card as a child process in its own process group, with a reduced environment that omits the model key. Host and connector exchange JSON lines over standard input and output through a small SDK, \code{ce.mjs}. Its MIDI calls clamp values to 0--127 and pass through the companion page's Web MIDI to USB instruments or to the virtual ports of the built-in synthesizers; the visual synthesizer is a full-screen WebGL water simulation on the laptop. Its gimbal calls go to a bridge that clamps pan to $\pm$120\textdegree, tilt to $\pm$45\textdegree{} and the rate to 10 commands/s, and its \code{mac} calls press keys through AppleScript System Events.

The tool \code{deploy\_connector} registers 1 to 16 actions with their ranges and waits 1.2\,s for output or a crash. The tool \code{test\_action} sends one value as an AR control would, waits 900\,ms, and returns the connector's log, separated into acknowledgements (lines beginning with $\rightarrow$) and refusals (\ding{55}). For MIDI and the gimbal, the host writes the acknowledgement line when the executor answers; for key presses, it relays the exit status of the connector's own \code{osascript} call. In either case, an acknowledgement establishes that a transport accepted a message, which is weaker than establishing that the device behaved as requested.

\subsection{Controls and Hand Programs}
\label{sec:hands}

A \emph{control} is an AR widget bound to one declared action. The agent may return up to six simple controls and an optional composed instrument of up to eight widgets, such as keys, pads, knobs and an XY surface; the prompt instructs it never to require pinch-and-drag gestures. A hand program is a JavaScript function \code{wire(s, out, mem)} of at most 4,096 characters that the headset calls on every frame. It reads per-hand signals (height, push and lateral position in a head-relative zone, smoothed with a One Euro filter \citep{casiez2012oneeuro}, plus pinch, fist, palm and speed) and, once a screen has been found, the fingertip's position on the screen and a touch flag. It acts through \code{out.set} and \code{out.trigger}.

The server rejects forbidden tokens such as imports, timers and \code{eval}, and runs the program on 480 synthetic frames in an empty \code{node:vm} context, requiring an average below 2\,ms per frame and valid, correctly scaled output. On the headset, the program runs in a Web Worker without network, storage or timer APIs, under a 4\,ms budget per call, a 50\,ms watchdog and a three-strike limit. An output gate passes only the actions in \code{wire.ready}, clamped to their ranges and limited to 30\,Hz. No model is called per hand frame (\dg{3}).

For touch on the laptop screen, the client estimates the screen's plane from the SAM bounding box and a prior on display size; the estimate can be off by more than 10\,cm. Screen signals remain null until a fingertip approaches the estimated plane and rests near it. The plane is then moved along the line of sight to pass through that fingertip.

\subsection{Safety and Stopping}
\label{sec:safety}

The \code{bash} tool passes through a deny-list of 20 regular expressions (covering \code{sudo}, any \code{kill}, the server's ports, credential files, \code{git push} and others), and the file tools write only inside the workspace; connectors themselves are not sandboxed (\S\ref{sec:discussion}). Stopping does not depend on the model (\dg{4}). The host sends \code{stop} to the connector, queues neutral releases (MIDI panic, sustain off, touched controllers set to zero, gimbal stop), and sends SIGTERM to the process group at 250\,ms and SIGKILL at 600\,ms; for 1.7\,s afterwards, only neutral releases from the retired connector pass. The wearer can stop generation on the card, stop a running program by holding both palms out for 400\,ms, or stop everything from the wrist menu, and a new request on the card retires the previous connector.

\subsection{Pose Recording}
\label{sec:recording}

Any session can be saved as a \emph{take}: a JSON-lines file that begins with approximate camera intrinsics and tracking metadata and continues with head and 25-joint hand poses at 30\,Hz, the protocol messages and input events, all timed from the start of the take. A hand that loses tracking is recorded as null and never as its last known pose. The Quest's system recording provides the video, and a review page plays the video and the poses on one timeline and exports the take.

\section{Technical Evaluation}
\label{sec:eval}

The technical evaluation characterizes the loop on real hardware through three questions. \eq{1}: Does the loop reach a ready state, with controls or a hand program bound to a running connector, for targets that use different transports? \eq{2}: How long does the wearer wait for the ready state, and where is that time spent? \eq{3}: What evidence does the system obtain that a connection affects the device?

\begin{table}[t]
\caption{The ten requests of the evaluation session, all addressed to the laptop's tag. Via: spoken request (voice) or suggestion chip. Returned: the output as logged; the log does not record whether a composed instrument was also returned. Ready: time to ready, excluding transcription. Calls: model calls, each returning one tool call.}
\label{tab:runs}
\centering
\footnotesize
\renewcommand{\arraystretch}{0.94}
\begin{tabular}{@{}lllllrr@{}}
\toprule
Run & Request (paraphrased) & Target & Via & Returned & Ready (s) & Calls \\
\midrule
R1  & Touch the visual synth with a finger  & Visual synth    & voice & hand program               & 31.6 & 9 \\
R2  & Scroll Reels by touch in the browser  & Reels in Chrome & voice & hand program               & 57.8 & 8 \\
R3  & Scroll Reels with gestures            & Reels in Chrome & voice & hand program + 3 controls  & 54.5 & 9 \\
R4  & Knobs for the OP-1 filter             & OP-1 Field      & voice & 2 controls                 & 24.9 & 7 \\
R5  & Control the OP-1 Field with fingers   & OP-1 Field      & voice & hand program               & 36.7 & 7 \\
R6  & ``Give me knobs for this''            & Laptop          & chip  & stopped by the wearer at 18.5\,s & n/a & 2 \\
R7  & Connect the gimbal to my fingers      & Osmo Mobile 4   & voice & hand program + 4 controls  & 30.0 & 6 \\
R8  & Controls for the water visualization  & Visual synth    & voice & 4 controls                 & 46.7 & 12 \\
R9  & Control the water with fingers        & Visual synth    & voice & hand program               & 32.5 & 6 \\
R10 & ``Give me knobs for this''            & Laptop          & chip  & 4 controls                 & 25.9 & 7 \\
\bottomrule
\end{tabular}
\end{table}

\subsection{Method}
\label{sec:method}

\paragraph{Apparatus.}
The client ran on a Meta Quest~3 in Quest Browser, and the server on a MacBook with the built-in agent engine and live Meta models. The targets were the built-in visual synthesizer, Instagram Reels in Chrome, the laptop itself, an OP-1 Field synthesizer connected over USB-MIDI, and a DJI Osmo Mobile~4 gimbal reached over Bluetooth through the Chrome bridge.

\paragraph{Procedure.}
We recorded one session on November 5, 2026 (06:21--07:07 JST), with the author as the only wearer. We label the requests R1--R10 in order (Table~\ref{tab:runs}). Requests were phrased freely rather than drawn from a fixed task set; eight were spoken and two used a suggestion chip. The server was restarted once, between R8 and R9. All ten requests addressed the laptop's tag, which had been created and named two days earlier, so the session contains no tagging or naming call. The wearer named the OP-1 and the gimbal in the request text (for example, ``this gimbal connected to my Mac''), so selecting those devices by pinch was not exercised. The agent's notes, which persist across sessions, already contained the OP-1's port name, its filter controller numbers and how to move the gimbal; R3 added the Reels keys. Several requests re-ask an earlier goal (R2 and R3, R4 and R5, R1 and R9), so the requests are not independent trials.

\paragraph{Measures.}
\emph{Time to ready} is the interval from the start of the server's handling of a \code{wire.request} to the corresponding \code{wire.ready}. It excludes speech transcription, which precedes the request and took 1.2--3.4\,s. For each request, we report the share of time to ready spent in model calls and the token counts, both from the per-call entries in the server log. Start and end times of individual calls were available only for R9 and R10. Gimbal acknowledgements were counted from the bridge's log lines. We did not measure headset frame time, the latency from hand movement to device response, the physical effect of any command, the error of the screen-plane estimate, or the detection accuracy of SAM~3.1.

\subsection{Results}
\label{sec:results}

\paragraph{Outcomes (\eq{1}).}
Nine of the ten requests reached a ready state, and none failed. The wearer stopped R6 after 18.5\,s; the log does not record the reason. Six runs returned a hand program and five returned controls, two of them both (Table~\ref{tab:runs}). The connectors were 15 to 42 lines long. All three transports were exercised: MIDI (visual synthesizer, OP-1), Bluetooth (gimbal) and key presses (Reels). R10's controls set the laptop's volume, scrolling, playback and mute; its connector made the volume calls through AppleScript directly, outside the three transports.

\paragraph{Time to ready (\eq{2}).}
Time to ready had a median of 32.5\,s (mean 37.8\,s, range 24.9--57.8\,s, $n = 9$). The ten runs used 73 model calls, with a median per-call latency of 3.48\,s (90th percentile 8.88\,s). Model calls accounted for a median of 89.9\% of each run's time to ready (range 82.5--96.2\%), and the remaining time had a median of 3.0\,s (range 2.1--5.5\,s). In R9 and R10, the remaining time consisted almost entirely of fixed waits, about 1.2\,s after \code{deploy\_connector} and 0.9\,s after each \code{test\_action}; the other tools took 1--6\,ms. A repeated request cost as much as the first, because every connector is written anew; R1 and R9, both requests to touch the water simulation, took 31.6\,s and 32.5\,s. Of 656,844 input tokens, 91.7\% were served from the prompt cache; from R2 onward, every run reused a 5,489-token prefix consisting of the system prompt and the tool schemas.

\paragraph{Evidence of effect (\eq{3}).}
The system's evidence of effect consists of acknowledgements. The gimbal bridge logged 889 rotate answers from the Osmo Mobile~4, all with status \code{00} (ok). Of these, 881 came from the agent's code (2 from \code{test\_action} and 879 while the wearer steered) and 8 from stop and release commands. At the bridge's limit of 10 commands/s, 879 moves imply at least 87.9\,s of steering. An answer indicates that the gimbal accepted a command; the logs contain no reliable angle. The author observed the ripples, the scrolling, the OP-1's sound and the gimbal's motion, which appear in the accompanying video, but the system measured none of these effects.

\paragraph{Pose recording and other model calls.}
The take covered R9 and R10. It lasted 111.6\,s, contained 3,247 pose frames at a mean of 30.01\,Hz, and ended when it reached the 24\,MiB cap. The left hand was untracked in 3.2\% of frames and the right hand in 2.6\%, including one interval of about 2.8\,s in which both were lost. SAM~3.1 answered 336 prompt calls with a median latency of 0.83\,s, and 40\% of the calls returned no object; the three frames with per-call records, each running three prompts in parallel, took 2.1--2.9\,s.

\section{Discussion}
\label{sec:discussion}

\paragraph{Integration at the time of use.}
In this session, device-specific knowledge entered the loop through the prompt's transport documentation and target hints, server-supplied context such as MIDI port names, and the agent's notes from earlier sessions. This suggests an architecture in which developers maintain a few documented transports and an agent writes the per-request integration code (\dg{1}), although outcomes then depend on accumulated state, such as the notes and the prompt cache, which we did not vary. The output form followed the request's wording: requests for knobs or controls returned controls only (R4, R8, R10), and those mentioning fingers, touch or gestures returned a hand program (R1--R3, R5, R7, R9), suggesting that the form can be left to the request; how wearers unaware of both forms would phrase requests is untested.

\paragraph{Generative UI for physical devices.}
Lindley et al. ask how HCI and design practice should evolve as AI models create interfaces \citep{lindley2026genui}. ControlAnything illustrates one division of labor for physical devices: developers prepare transports, checks and stop paths, and the model generates each interface together with its device code. A generated interface for a physical device carries a risk that a generated web page does not, because the device may not behave as the widgets suggest. The server's checks ensure that every widget binds an action of a running connector, not that the action has its intended physical effect.

\paragraph{The latency budget.}
Half a minute is long for an interactive system, but it is a setup cost paid once per request; afterwards, the mapping runs on every frame without model calls (\dg{3}). With about 90\% of the setup time spent in sequential model calls, reductions must come from fewer or faster model calls, or from reuse. Reusing a connector already tested on the same device could avoid most of the cost of repeated requests such as R1 and R9, at the price of the guarantee that the code was written for the current request. We did not measure the latency of the running control loop; musical control in particular demands low and consistent latency \citep{wessel2002problems}, so this measurement must precede any claim of instrument-grade control.

\paragraph{From acknowledgement to physical effect.}
An acknowledgement establishes that a transport accepted a message, not that the device did what the wearer asked. The agent's notes from the day before the session record OP-1 MIDI messages that were acknowledged while the instrument remained silent, until its MIDI receive setting was turned on. Because \code{test\_action} classifies log lines by their first character, a connector could also print an acknowledgement itself. AutoBridge closes this gap with a human observer \citep{liu2025autobridge}, and LoopHIL with laboratory instruments, observing that smaller models often report unverified claims \citep{shah2026loophil}. In MR, the headset's own camera and microphone could observe many such effects (ripples on a screen, a moving gimbal, a sound) without a separate observer.

\paragraph{Safety of runtime-generated code.}
Connectors are ordinary Node processes with the laptop user's permissions and are not sandboxed; the deny-list does not inspect their code, and the project's documentation states that it is not a sandbox. Headset code, in contrast, is dry-run and confined to a restricted Worker. The SDK limits what a connector sends through it (clamped MIDI values, no SysEx, a gimbal envelope enforced in the bridge), but a connector is arbitrary Node and can bypass it, and prompt injection through camera-visible text or device output is handled by a prompt rule alone. Runtime-generated device code therefore needs operating-system containment, scoped to the transports a request uses, before use outside a controlled setting.

\paragraph{Limitations and future work.}
\label{sec:limits}
The evaluation has one wearer (the author), one session, one laptop and one tag; its requests are not independent trials, pinch selection of the OP-1 and the gimbal was not exercised, and a request addresses one card, so rules spanning several devices are out of reach. Apart from R10's volume calls, every target was reached through the three hand-written transports, so the evaluation does not show that the agent can integrate a device that has no transport. This paper reports a technical evaluation rather than a user study, so we make no claims about learnability, comfort, ease of use or naturalness; a study with first-time wearers is planned to examine them. Accessibility was not evaluated either; the design relies on two tracked hands for the palms-out stop, the left hand for the wrist menu, fine pinches and intelligible speech. The quantities listed as unmeasured in \S\ref{sec:method} remain open. Future work follows from these limits: tags on the devices themselves, an operating-system sandbox for connectors, effect tests through the headset's camera or microphone, connector reuse and hand-to-device latency measurement.

\section{Conclusion}

ControlAnything treats both the interface to a nearby device and the connection beneath it as generated artifacts that a coding agent writes on request, on top of hand-written transports. In one session on a real headset, it returned controls or hand programs for five targets with a median time to ready of 32.5\,s, most of it spent in model calls; its evidence of success ended at transport acknowledgements. Shortening this wait and closing the gap to physical effect are the main open problems.

\section*{Acknowledgments}
This project was built for the Meta Global AI Developer Hackathon (Indie track), November 2026. This work is not affiliated with or endorsed by Meta. The software and this manuscript were written with the help of AI coding and writing assistants; the author checked every claim and number against the session logs and the code.

\bibliographystyle{unsrtnat}
\bibliography{refs}

\end{document}